%% file: dstotaunu.tex
\RequirePackage[mathlines]{lineno}
\documentclass[twocolumn,showpacs,aps,prl]{revtex4-2}
\usepackage{overpic,graphicx}
\usepackage{dcolumn}
\usepackage{bm}
\usepackage{rotating}
\usepackage{subfigure}
\usepackage{color}
\usepackage[dvipdfmx, bookmarksnumbered, pdfstartview=FitH,colorlinks,urlcolor=blue, citecolor=blue,linkcolor=blue,] {hyperref}
\usepackage{lineno}
\usepackage{multirow}
\usepackage{amsmath}
\usepackage{makecell}%2020.12.10, make the text to be center/right/left

\newcommand{\dstoksk}{D_{s}^{-} \to K_{S}^{0}K^{-}}

\newcommand{\ksk}{K_{S}^{0}K^{-}}
\newcommand{\kkpi}{K^{+}K^{-}\pi^{-}}
\newcommand{\kkpipiz}{K^{+}K^{-}\pi^{-}\pi^{0}}
\newcommand{\kskpipi}{K_{S}^{0}K^{-}\pi^{+}\pi^{-}}
\newcommand{\kskpipim}{K_{S}^{0}K^{+}\pi^{-}\pi^{-}}
\newcommand{\pipipi}{\pi^{+}\pi^{-}\pi^{-}}
\newcommand{\pieta}{\pi^{-}\eta}
\newcommand{\pipizeta}{\pi^{-}\pi^{0}\eta}
\newcommand{\pietapgam}{\pi^{-}\eta'_{\pi^{+}\pi^{-}\eta}}
\newcommand{\pietaprho}{\pi^{-}\eta'_{\gamma \rho^{0}}}
\newcommand{\kpipi}{K^{-}\pi^{+}\pi^{-}}

\newcommand{\etot}{E_{\mathrm{extra}}^{\mathrm{tot}}}

\begin{document}

%\linenumbers

\title{\bf \boldmath
Measurement of the Absolute Branching Fraction of $D_s^+ \to \tau^+ \nu_{\tau}$ via $\tau^+ \to e^+ \nu_e \bar{\nu}_{\tau}$
}

\author{\input{author}}

\begin{abstract}
Using a dataset of 6.32 fb$^{-1}$ of $e^+ e^-$ annihilation data collected with the BESIII detector at center-of-mass energies between 4178 and 4226 MeV, we have measured the absolute branching fraction of the leptonic decay $D_s^+ \to \tau^+ \nu_{\tau}$ via $\tau^+ \to e^+ \nu_e \bar{\nu}_{\tau}$, and find $\mathcal{B}_{D_s^+ \to \tau^+ \nu_{\tau}}=(5.27\pm0.10\pm0.12)\times10^{-2}$, where the first uncertainty is statistical and the second is systematic. The precision is improved by a factor of 2 compared to the previous best measurement. Combining with $f_{D_s^+}$ from Lattice quantum chromodynamics calculations or the $|V_{cs}|$ from the CKMfitter group, 
we extract $|V_{cs}|=0.978\pm0.009\pm0.012$ and $f_{D_s^+}= (251.1\pm2.4\pm3.0)$ MeV, respectively. Combining our result with the world averages of $\mathcal{B}_{D_s^+ \to \tau^+ \nu_{\tau}}$ and $\mathcal{B}_{D_s^+ \to \mu^+ \nu_{\mu}}$, we obtain the ratio of the branching fractions $\mathcal{B}_{D_s^+ \to \tau^+ \nu_{\tau}} / \mathcal{B}_{D_s^+ \to \mu^+ \nu_{\mu}} = 9.72\pm0.37$, which is consistent with the standard model prediction of lepton flavor universality.  

\end{abstract}


\maketitle

\oddsidemargin  -0.2cm
\evensidemargin -0.2cm

%-------Introduction----------
Leptonic decays of charged pseudoscalar mesons can provide accurate determinations of Cabibbo-Kobayashi-Maskawa (CKM) matrix elements and a clean setting for tests of the lepton flavor universality (LFU). In the standard model (SM), the partial decay width of $D_s^+ \to \ell^+ \nu_{\ell}$ ($\ell~=~e,~\mu,~\tau$) to the lowest order is given by~\cite{decayratedstotaunu}
\begin{small}
\begin{eqnarray}
  \label{decayratedstolnu_draft}
  \Gamma_{D_{s}^{+} \to \ell^+ \nu_\ell} =\frac{G_{F}^{2} f^{2}_{D_s^+} m_{D_s^+}^3}{8\pi} |V_{cs}|^{2} \mu_\ell^2 ( 1 - \mu_\ell^2)^{2},
\end{eqnarray}
\end{small}%
where $G_{F}$ is the Fermi coupling constant, $f_{D_s^+}$ is the decay constant parameterizing strong-interaction effects between the two initial-state quarks, 
$V_{cs}$ is the $c \to s$ CKM matrix element, and $\mu_{\ell}$ is the ratio of the $\ell^+$ lepton mass to the $D_s^+$ meson mass, $m_{D_s^+}$. 

From Eq.~\eqref{decayratedstolnu_draft}, it is clear that the ratio of the decay widths $\Gamma_{D_s^+ \to \tau^+ \nu_{\tau}}/\Gamma_{D_s^+ \to \mu^+ \nu_{\mu}}$ only depends on $\mu_{\tau}$ and $\mu_{\mu}$, and is equal to $9.75\pm0.01$~\cite{pdg}. This is consistent with the experimental measurements~\cite{pdg} given the current relatively large uncertainties. In recent years, some hints of LFU violation in semileptonic $B$ decays have been reported by different experiments~\cite{flv1,flv2,flv3,flv4,flv5}. It is argued that the violation may occur in $c \to s$ transitions due to the interference between different amplitudes~\cite{lfu_dstotaunu} or the interactions with scalar operators~\cite{liying}. Therefore, the improved precision on the measurements of $D_s^+ \to \ell^+ \nu_\ell$ will be of great interest in testing LFU.

Given $f_{D_s^+}$ from Lattice quantum chromodynamics (LQCD) calculations~\cite{fds_flag}, a precision measurement of the branching fraction (BF) of $D_s^+ \to \ell^+ \nu_\ell$ ($\mathcal{B}_{D_s^+ \to \ell^+ \nu_\ell}$) allows an accurate determination of $|V_{cs}|$ and a stringent test of the unitarity of the CKM matrix. Conversely, given $|V_{cs}|$ from the SM global fit~\cite{pdg}, the $\mathcal{B}_{D_s^+ \to \ell^+ \nu_\ell}$ value allows one to extract $f_{D_s^+}$ and verify the theoretical determinations of the decay constant.

Previous measurements of $\mathcal{B}_{D_s^+ \to \tau^+ \nu_{\tau}}$ from CLEO~\cite{cleo_dstotaunu_enunu1,cleo_dstotaunu_enunu2,cleodstotaunu2,cleodstotaunu3}, $BABAR$~\cite{babardstotaunu}, Belle~\cite{belldstotaunu}, and BESIII~\cite{besiiidstotaunu, dstotaunu_hajime, dstotaunu_panxiang} have limited precision. In this Letter, we report an improved measurement of $\mathcal{B}_{D_s^+ \to \tau^+ \nu_{\tau}}$ using the $\tau^+ \to e^+ \nu_e \bar{\nu}_{\tau}$ decay~\cite{charge_conjugate} with precision improved by a factor of 2 compared to the previous best measurement~\cite{dstotaunu_hajime}. The data samples used, corresponding to a total integrated luminosity of 6.32 fb$^{-1}$, were collected at center-of-mass energies ($E_{\rm cm}$) of 4178, 4189, 4199, 4209, 4219, and around 4226 MeV~\cite{ref_emc_4230_hajime, ref_lumi_4230_hajime} with the BESIII detector~\cite{Ablikim:2009aa,luopw_ds} operating at the BEPCII collider~\cite{Yu:IPAC2016-TUYA01}.

Simulated data samples produced with a {\sc geant4}-based~\cite{geant4} Monte Carlo (MC) package, which incorporates the geometric description of the BESIII detector and the detector response, are used to determine detection efficiencies and to estimate backgrounds. The MC simulation includes the beam-energy spread and initial-state radiation (ISR). Moreover, a generic MC sample corresponding to 40 times the integrated luminosity of data is produced, including the production of open-charm final states, ISR production of vector charmonium states, and continuum processes. Details can be found in Ref.~\cite{dstotaunu_hajime}.
   
%---------Ananlysis method-----------
We apply a ``double-tag" (DT) technique~\cite{dtmethod1, dtmethod2} to select the leptonic $D_s^+$ decays. Events of $e^+ e^- \to D_s^{*+}(\to\gamma /\pi^0 D_s^{+})D_s^{-}$ and $e^+ e^- \to (\gamma_{\rm ISR})D_s^+ D_s^-$ are identified by fully reconstructing a hadronic $D_s^-$ decay (``single tag" or ``ST") from eleven decay modes: $\ksk$, $\kkpi$, $\kkpipiz$, $\kskpipi$, $\kskpipim$, $\pipipi$, $\pieta$, $\pipizeta$, $\pietapgam$, $\pietaprho$, and $\kpipi$, of which intermediate mesons are reconstructed by $K_S^0 \to \pi^+ \pi^-$, $\pi^0 (\eta)\to \gamma \gamma$, $\eta'_{\pi^+\pi^-\eta}\to\pi^+\pi^-\eta$, $\eta'_{\gamma \rho^0}\to\gamma \rho^0$, and $\rho^0\to\pi^+ \pi^-$, respectively. A DT signal event consists of an ST $D_s^-$ candidate accompanied by a $D_s^+ \to \tau^+ \nu_{\tau}$ signal candidate.

%-----------how to obtain the branching fraction-----------
The absolute $\mathcal{B}_{D_s^+ \to \tau^+ \nu_{\tau}}$ is given by
\begin{eqnarray}
  \label{dtbr}
  \mathcal{B}_{D_s^+ \to \tau^+ \nu_{\tau}} = \frac{N_{\rm DT} / \epsilon_{\rm DT}}{ \sum_{i} (N_{\rm ST}^{i}/\epsilon_{\rm ST}^{i}) \mathcal{B}_{\tau^+ \to e^+ \nu_e \bar{\nu}_{\tau} }},
\end{eqnarray}
where $i$ indicates the data samples at the six energy points, $N_{\rm ST}^{i}$ ($N_{\rm DT}$) and $\epsilon_{\rm ST}^{i}$ ($\epsilon_{\rm DT}$) are the ST (DT) yields and efficiencies, and $\mathcal{B}_{\tau^+ \to e^+ \nu_e \bar{\nu}_{\tau}}$ is the world average BF of $\tau^+ \to e^+ \nu_e \bar{\nu}_{\tau}$~\cite{pdg}. The ST yields $N_{\rm ST}^i$ are obtained for each energy point, while the DT yield $N_{\rm DT}$ is obtained for the combined data sample because of limited statistics. 

%-------Event selections for ST candidates--------------
The ST $D_s^-$ candidates are reconstructed from the above eleven hadronic decay modes with almost the same selection criteria as those in Ref.~\cite{dstomunu_zhangsf}, the differences are described here. For $D_s^- \to K_S^0 K^-$, the $K_S^0$ candidate is not required to have a decay length larger than twice the vertex resolution. For $D_s^- \to \pi^+ \pi^- \pi^-$, the $K_S^0$ candidates are retained in the invariant mass of either $\pi^+ \pi^-$ combination. The specific ionization energy loss ($dE/dx$) in the main drift chamber (MDC) and the time-of-flight (TOF) information are combined and used for particle identification (PID) by forming confidence levels for the charged pion and kaon hypotheses ($CL_{\pi}$, $CL_{K}$). Kaons are identified by requiring $CL_K > CL_{\pi}$, while pions except for those from $K_S^0$ decays are demanded to satisfy $CL_{\pi}>CL_K$. To suppress background from Bhabha events, the relative-probability sum, $\sum_{j=1}^{n} [(CL^j_e)/(CL^j_{\pi}+CL^j_K+CL^j_e)]$, must be smaller than $2.0$ ($0.9$) for $n>1$ ($n=1$), where $n$ is the number of charged tracks for the ST mode.

The recoil mass against the tag $D_s^-$, $M_{\rm rec}=\sqrt{(E_{\rm cm} - \sqrt{|\vec{p}_{D_s^-}|^2+m_{D_s^-}^2})^2-|\vec{p}_{D_s^-}|^2}$, is used to select the $e^+ e^- \to (\gamma_{\rm ISR}) D_s^{(*)\pm} D_s^{\mp}$ events, where $\vec{p}_{D_s^-}$ is the momentum of the reconstructed $D_s^-$ and $m_{D_s^-}$ is the $D_s^-$ nominal mass. Events with $M_{\rm rec}$ lying within the mass windows of [2.050, 2.195], [2.048, 2.205], [2.046, 2.215], [2.044, 2.225], [2.042, 2.235], and [2.040, 2.220] GeV/$c^2$~\cite{dstotaunu_hajime, mass_window} are retained for further analysis for the data samples at $E_{\rm cm}=$ 4178, 4189, 4199, 4209, 4219, and around 4226 MeV, respectively. In each event, we only keep the tag $D_s^-$ candidate with the $M_{\rm rec}$ closest to the nominal $D_s^{*}$ mass~\cite{pdg} per tag mode per charge.

\begin{figure}[hbpt]
\centering
   \includegraphics[width=0.50\textwidth]{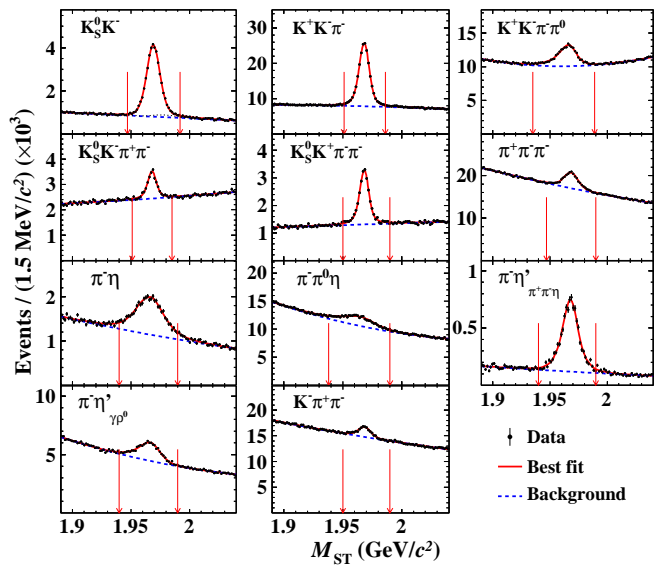}
\caption{Fits to the $M_{\rm ST}$ distributions of the ST $D_s^-$ candidates for the data sample at $E_{\rm cm}=4178$ MeV. The difference between the dotted black line and the dashed blue line for $D_s^- \to K_S^0 K^-$ is the $D^-\to K_S^0 \pi^-$ background. The pairs of arrows denote the signal regions.}
%  \caption{Fits to the $M_{\rm ST}$ distributions of the ST $D_s^-$ candidates for the data sample at $E_{\rm cm}=4178$ MeV. Dots with error bars are data, solid red lines are the total fits, dashed blue lines are the combinatorial backgrounds, and the dotted black line for $D_s^- \to K_S^0 K^-$ is from the $D^-\to K_S^0 \pi^-$ background. The pairs of arrows denote the signal regions.}
\label{ST_yield_data_4178}
\end{figure}

The ST yields are obtained from fits to the invariant mass spectra of the ST $D_s^-$ candidates ($M_{\rm ST}$). In the fits, the $D_s^-$ signal is modeled by the MC-simulated shape convolved with a Gaussian resolution function. The combinatorial background is described by a first to third order Chebychev function. For $D_s^- \to K_S^0 K^-$, the peaking background from $D^- \to K_S^0 \pi^-$ is incorporated with the MC-simulated shape and its yield is left free. The fit results for the data sample at $E_{\rm cm}=4178$ MeV in the range of $M_{\rm ST}\in [1.89, 2.04]$ GeV/$c^2$ are shown in Fig.~\ref{ST_yield_data_4178} as an example. For each tag mode, the ST yield is obtained within a tag-dependent $M_{\rm ST}$ window corresponding to approximately $\pm 3\sigma$ of the $D_s$ mass, where $\sigma$ is the fitted resolution of $M_{\rm ST}$. Following the same procedure, the ST efficiencies are estimated using the generic MC sample. Table~\ref{dtyields_efficiency} lists the sum ($R$) of the ratios of the ST yield over the ST efficiency from the six data samples.

%2021.8.21,
%/scratchfs/bes/hjli/Ds4180/703/best_ST_XYZ/data/total_rn/init_dt/com_6_eneries/draft/STyield4180data/hajime/PRL_published/fit_ST_yield_40x_data.eps

\begin{table*}[htbp]
\begin{center}
\caption{The ratios of the ST yield in data over the ST efficiency summed over the six data samples ($R=\sum_i (N_{\rm ST}^{i}/\epsilon_{\rm ST}^{i})$), the numbers of DT events ($N^{\rm tot}_{\rm DT}$), the numbers of backgrounds of non-$D_s^-$, $D_s^+ \to K^0_L e^+ \nu_e$, and $D_s^+ \to X e^+ \nu_e$ in the $\etot$ (defined in the signal $D_s^+$ analysis part) signal region ($N^{{\rm non}{\text -}D_s^-}_{\rm DT}$, $N^{K^0_L e^+ \nu_e}_{\rm DT}$, and $N_{\rm DT}^{X e^+ \nu_e}$), the numbers of DT events after removing backgrounds ($N_{\rm DT}$), the DT efficiencies ($\epsilon_{\rm DT}$), and the obtained $\mathcal{B}_{D_s^+ \to \tau^+ \nu_{\tau}}$. Both $\epsilon_{\rm ST}^{i}$ in $R$ and $\epsilon_{\rm DT}$ include the relevant BFs for $K_S^0$, $\pi^0$, $\eta$, $\eta'$, and $\rho^0$ decays. The $N^{K^0_L e^+ \nu_e}_{\rm DT}$ is fixed in the fit and its uncertainty will be treated later as a systematic uncertainty. For $\mathcal{B}_{D_s^+ \to \tau^+ \nu_{\tau}}$, the first, second, and third uncertainties are statistical, tag-mode dependent systematic and tag-mode independent systematic, respectively. The statistical uncertainty from the ST yield in $R$ is included in the second uncertainty of $\mathcal{B}_{D_s^+ \to \tau^+ \nu_{\tau}}$. For the other numbers, the uncertainties are statistical only.}
\label{dtyields_efficiency}
\setlength{\extrarowheight}{0.2ex}
 \setlength{\tabcolsep}{3.5pt}
  \renewcommand{\arraystretch}{1.0}
\begin{tabular}{l  c   r@{$\,\pm\,$}l  r@{$\,\pm\,$}l  c  r@{$\,\pm\,$}l  r@{$\,\pm\,$}l  r@{$\,\pm\,$}l  l@{$\,\pm\,$}r @{$\,\pm\,$}r @{$\,\pm\,$}r}\hline\hline
Mode &$R(\times10^{4})$ & \multicolumn{2}{c}{$N_{\rm DT}^{\rm tot}$}  & \multicolumn{2}{c}{$N^{{\rm non}\textnormal{-}D_s^-}_{\rm DT}$} & $N^{K^0_L e^+ \nu_e}_{\rm DT}$ & \multicolumn{2}{c}{$N^{X e^+ \nu_e}_{\rm DT}$} & \multicolumn{2}{c}{$N_{\rm DT}$} & \multicolumn{2}{c}{$\epsilon_{\rm DT}\,(\%)$ }& \multicolumn{4}{c}{$\mathcal{B}_{D_s^+ \to \tau^+ \nu_{\tau}}\,(\%)$  }\\ \hline
$\ksk$      	& 	17.5			& 	500	& 	22	& 	30.2	& 	2.9	& 	60	& 	56.9	& 	2.7	& 	353	& 	23	& 	23.69	& 	0.22	& 	4.79	& 	0.31	& 	0.07	& 	0.11	\\
$\kkpi$     	& 	62.7			& 	2394	& 	49	& 	544.5	& 	15.0	& 	154	& 	202.8	& 	5.0	& 	1493	& 	51	& 	26.63	& 	0.13	& 	5.03	& 	0.17	& 	0.04	& 	0.11	\\
$\kkpipiz$  	& 	69.5			& 	1198	& 	35	& 	386.1	& 	9.5	& 	55	& 	92.2	& 	5.5	& 	664	& 	36	& 	9.52	& 	0.08	& 	5.64	& 	0.31	& 	0.17	& 	0.12	\\
$\kskpipi$  	& 	11.3			& 	203	& 	14	& 	82.2	& 	5.1	& 	9	& 	14.8	& 	2.0	& 	97	& 	15	& 	8.51	& 	0.22	& 	5.68	& 	0.90	& 	0.23	& 	0.13	\\
$\kskpipim$ 	& 	18.5			& 	291	& 	17	& 	59.1	& 	4.4	& 	16	& 	26.6	& 	2.4	& 	189	& 	18	& 	9.95	& 	0.15	& 	5.78	& 	0.54	& 	0.13	& 	0.13	\\
$\pipipi$   	& 	13.7			& 	952	& 	31	& 	323.3	& 	12.0	& 	49	& 	70.7	& 	3.8	& 	509	& 	33	& 	38.39	& 	0.37	& 	5.43	& 	0.36	& 	0.13	& 	0.12	\\
$\pieta$    	& 	19.1			& 	359	& 	19	& 	32.1	& 	3.8	& 	23	& 	40.4	& 	2.1	& 	264	& 	19	& 	12.93	& 	0.15	& 	6.00	& 	0.44	& 	0.22	& 	0.13	\\
$\pipizeta$ 	& 	105.9			& 	1065	& 	33	& 	229.1	& 	7.9	& 	65	& 	100.8	& 	5.2	& 	670	& 	34	& 	6.65	& 	0.05	& 	5.35	& 	0.27	& 	0.19	& 	0.12	\\
$\pietapgam$	& 	43.6			& 	167	& 	13	& 	1.0	& 	0.3	& 	11	& 	17.7	& 	1.4	& 	137	& 	13	& 	2.56	& 	0.04	& 	6.93	& 	0.66	& 	0.17	& 	0.15	\\
$\pietaprho$	& 	47.3			& 	478	& 	22	& 	92.0	& 	5.6	& 	32	& 	53.3	& 	3.2	& 	301	& 	23	& 	7.21	& 	0.08	& 	4.96	& 	0.38	& 	0.14	& 	0.11	\\
$\kpipi$    	& 	7.2			& 	787	& 	28	& 	466.3	& 	14.6	& 	22	& 	35.2	& 	3.5	& 	263	& 	32	& 	34.45	& 	0.69	& 	5.96	& 	0.72	& 	0.25	& 	0.13	\\
\hline\hline
\end{tabular}
\vspace{-0.1cm}
\end{center}
\end{table*}

%-------DT analysis--------------------
In the presence of the ST $D_s^-$ candidate, we select the signal candidate for $D_s^+ \to \tau^+ \nu_{\tau}$ with $\tau^+ \to e^+ \nu_e \bar{\nu}_{\tau}$. Given the very small variation of detection efficiencies and backgrounds at different energy points, the six data samples are combined for further analysis. The DT candidates are required to have exactly one charged track in addition to the daughters of the tag side and that track must have a charge opposite that of the tag side decay. The track is also required to be associated with a good electromagnetic calorimeter (EMC) shower, as described in Ref.~\cite{dstomunu_zhangsf}. To identify the positron, the combined $dE/dx$, TOF, and EMC information is used to determine a $CL_e$. We assign the track as a positron if it satisfies $CL_e>0.1\%$ and $CL_e/(CL_e+CL_{\pi}+CL_K)>0.8$. The candidate track is further required to have a momentum in the MDC greater than 0.2 GeV/$c$ and a ratio of the energy deposited in the EMC to the momentum greater than 0.8.

\begin{figure}[htbp]
\centering
   \includegraphics[width=0.50\textwidth]{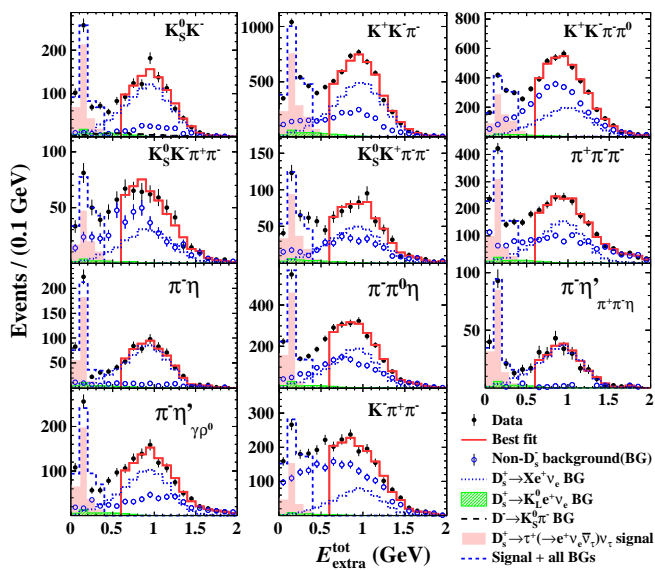}
  \caption{The $\etot$ distributions of the DT candidates. The signal component is normalized to our measured BF.} 
  %\caption{The $\etot$ distributions of the DT candidates. The signal is normalized to our measured branching fraction, and the signal plus all backgrounds only lies in $\etot<0.4$.} 
  %\caption{The $\etot$ distributions of the DT candidates in the combined data sample, along with results of fits to $\etot>0.6$ GeV, background estimates for non-$D_s^-$, $D_s^+ \to K_L^0 e^+ \nu_e$, $D_s^+ \to X e^+ \nu_e$, and $D^-\to K_S^0 \pi^-$ only for $\dstoksk$, the expected signal $D_s^+ \to \tau^+ (e^+ \nu_e \bar{\nu}_{\tau}) \nu_{\tau}$, and the expected signal plus all backgrounds in $\etot<0.4$.} 
\label{fit_etot_data_4178}
\end{figure}

To effectively discriminate signal from background, we adopt the variable $\etot$ following Refs.~\cite{cleo_dstotaunu_enunu1,cleo_dstotaunu_enunu2}. It is the total energy of the good EMC showers~\cite{goodshower}, excluding those associated with the ST selection and those within $5^\circ$ of the initial direction of the positron. The latter eliminates energy associated with final state radiation (FSR) from the positron track. We do not exclude soft photons or $\pi^0$s originating directly from the $D_s^{*}$. The $\etot$ distributions of the DT candidates for various tag modes in the combined data sample are shown in Fig.~\ref{fit_etot_data_4178}. Entries in the lowest bin include those events without any extra good EMC showers. We determine the background yield by a fit to the region $\etot>0.6$ GeV, where the signal is negligible, and extrapolate the backgrounds into the signal region using MC-derived shapes. The signal yield is then determined in the region $\etot<0.4$ GeV by statistically subtracting the expected backgrounds from the DT events seen in data. This procedure is insensitive to the signal shape, except for the inefficiency introduced by the definition of the signal region.

The backgrounds in the $\etot$ distributions can be divided into three categories. The first one is the non-$D_s^-$ background with an incorrectly reconstructed ST $D_s^-$. The second is the $D_s^+ \to K_L^0 e^+ \nu_e$ background, which survives when the $K_L^0$ passes through the detector without decaying or significantly interacting. The third is the $D_s^+ \to Xe^+ \nu_e$ background, which is dominated by the six semileptonic decays $D_s^+ \to \eta e^+ \nu_e$, $\eta' e^+ \nu_e$, $\phi e^+ \nu_e$, $f_0(980) e^+ \nu_e$, $K^{*}(892)^{0} e^+ \nu_e$, and $K_S^0 e^+ \nu_e$. The latter two cases are dominated by correctly reconstructed ST $D_s^-$.

Binned maximum likelihood fits are performed in the region $\etot>0.6$ GeV. 
The shape and size of the non-$D_s^-$ background are determined from the events in the $M_{\rm ST}$ sideband regions ([1.895, 1.92] and [2.01, 2.035] GeV/$c^2$). 
For the tag modes with neutral daughters, the resolution difference between data and MC simulation (called data-MC difference) has been corrected. 
The shape of the $D_s^+ \to K_L^0 e^+ \nu_e$ background is modeled by the MC-derived shape corrected by a two dimensional (polar angle and momentum) data-MC difference for the $K_L^0$ detector response. These correction factors are obtained by using a control sample of $D^0 \to K_L^0 \pi^+ \pi^-$ from 2.93 fb$^{-1}$ of data collected at $E_{\rm cm}=$ 3.773 GeV~\cite{luminosity_3773}. The background yield is calculated with $\mathcal{B}_{D_s^+ \to K^0 e^+ \nu_e}=(3.25\pm0.38\pm0.16)\times10^{-3}$, quoted from our previous work~\cite{lilei_paper_Ds_k0enu}. The peaking background from $D^- \to K_S^0 \pi^-$ is present only for the $\dstoksk$ tag mode and its yield is estimated from the $M_{\rm ST}$ fit. 
The yield of the $D_s^+ \to X e^+ \nu_e$ background is left free, with the shape extracted from the MC simulation with the individual BFs for the six background channels fixed as $\mathcal{B}_{D_s^+ \to \eta e^+ \nu_e} = (2.32\pm 0.08)\%$~\cite{phienu_3, etaenu_1, etaenu_2}, $\mathcal{B}_{D_s^+ \to \eta' e^+ \nu_e} = (0.80\pm 0.07)\%$~\cite{phienu_3, etaenu_1, etaenu_2}, $\mathcal{B}_{D_s^+ \to \phi e^+ \nu_e} = (2.37\pm 0.11)\%$~\cite{phienu_2, phienu_3, phienu_4}, $\mathcal{B}_{D_s^+ \to f_0 (980) e^+ \nu_e,~f_0(980)\to \pi \pi} = (0.30\pm 0.05)\%$~\cite{f0enu_1, alex_1128_2018_BESIII_Collaboration}, $\mathcal{B}_{D_s^+ \to K^{*}(892)^0 e^+ \nu_e} = (0.21\pm 0.03)\%$~\cite{lilei_paper_Ds_k0enu, phienu_3}, and $\mathcal{B}_{D_s^+ \to K^0 e^+ \nu_e} = (0.34\pm 0.04)\%$~\cite{lilei_paper_Ds_k0enu,phienu_3}. 
Moreover, the MC-based shapes have been further weighted by the individual ST yields at various energy points. 

Table~\ref{dtyields_efficiency} lists the obtained DT yields and DT efficiencies, where the latter are evaluated from the generic MC sample.

In the BF measurement, most uncertainties related to the ST selection are cancelled. The remaining systematic uncertainties are divided into two cases. The first case is from tag-mode dependent systematic uncertainties. 

Systematic uncertainties in the ST yield are examined by changing the fit range, the signal and background shapes, and the bin size and the background fluctuation of the fitted ST yield. The alternative fit range is chosen as [1.895, 2.035] GeV/$c^2$, corresponding to approximately $1\sigma$ of $M_{\rm ST}$ reduced from both sides of the nominal range. The nominal signal shapes obtained from the generic MC sample are replaced with those from the signal MC sample. The background shape is changed to a different order of the Chebychev function. The bin size is doubled or halved. For each variation, the efficiency-corrected yields are found to be consistent. The differences in the ratio of the ST yield over the ST efficiency for a given ST mode for all variations, and the background fluctuation of the fitted ST yield, are weighted by the ST yields in various data samples, and added in quadrature. The resulting overall systematic uncertainty of the ST yield is 0.61\%.

Tag bias related to the ST selection arises from different event environments (e.g., charged and neutral multiplicities). The ratios of the ST efficiencies from the generic MC sample and the signal MC sample for various tag modes are examined. The difference between the two ST efficiencies is weighted by the ST yields in various data samples. The systematic uncertainty due to tag bias is assigned to be 0.26\%.

For the non-$D_s^-$ background, we replace the distributions from the $M_{\rm ST}$ sidebands with those for the background events in the $M_{\rm ST}$ signal region from the generic MC sample. The change of the measured BF, 0.07\%, is assigned as the systematic uncertainty. 
In the nominal fit, the MC-based correction signal events in the $M_{\rm ST}$ sideband regions are normalized using the ratio of signal yields in the $M_{\rm ST}$ distributions of data and the generic MC sample. Alternative fits are performed $10^{4}$ times with the ratio varied according to a random Gaussian sampling based on its statistical error. The distribution of the relative difference on the DT yield is fitted by a Gaussian function, and the width of 0.07\% is taken as the systematic uncertainty. Here, and below where this method is also used, no significant biases in the means are observed. 
The systematic uncertainty from the $D^- \to K_S^0 \pi^-$ background for the $D_s^- \to K_S^0 K^-$ tag mode is assigned to be 0.05\% with a similar technique. The uncertainty from the limited MC statistics is 0.31\%. The quadratic sum of all the tag-mode dependent systematic uncertainties is 0.74\%.

The second case is from tag-mode independent systematic uncertainties. The systematic uncertainty associated with the $\etot<0.4$ GeV signal region is estimated using the DT events of $D_s^+ \to \pi^+ (\pi^0) \eta$ as the control sample. MC studies have verified that the efficiency for the requirement of $\etot<0.4$ GeV in the control sample is consistent with that in our signal decay. The systematic uncertainty is assigned to be 1.0\% considering the data-MC difference of the efficiencies.

The systematic uncertainty related to extra charged tracks is studied with the DT events of $D_s^+ \to \pi^+ \phi(\to K^+K^-)$ and $D_s^+ \to K^+ \bar{K}^{*}(892)^0(\to K^-\pi^+)$. The data-MC difference of the efficiencies, 1.4\%, is conservatively taken as the corresponding systematic uncertainty.

The systematic uncertainties associated with the $e^+$ tracking~\cite{alex_1128_2018_BESIII_Collaboration} and PID efficiencies are studied with a control sample of radiative Bhabha events. The two dimensional (momentum and the polar angle) efficiencies in the control sample are re-weighted to match those in our signal decay. The data-MC difference on $e^+$ tracking (PID) efficiency, 0.3\% (0.3\%), is assigned as the relevant systematic uncertainty.

The systematic uncertainty from the shape of $D_s^+ \to K_L^0 e^+ \nu_{e}$ background is estimated by replacing the corrected shape of $\etot$ with the uncorrected one. The change of the measured BF is negligible. We also examine the size of the $D_s^+ \to K_L^0 e^+ \nu_{e}$ background by sampling the quoted BF~\cite{lilei_paper_Ds_k0enu} 10$^4$ times based on a Gaussian distribution given by its uncertainty. The width of the relative difference of the fitted DT yield, 1.2\%, is assigned as the systematic uncertainty.

The uncertainty coming from the fixed $D_s^+ \to Xe^+ \nu_e$ background shape is examined by varying the proportion of each of the six main background modes via sampling their BFs~\cite{lilei_paper_Ds_k0enu, etaenu_1, etaenu_2, phienu_3, phienu_2, phienu_4, f0enu_1, alex_1128_2018_BESIII_Collaboration} $10^4$ times, based on Gaussian distributions given by their uncertainties. The width of the relative difference of the DT efficiency, weighted by the ST yields in various data samples, 0.1\%, is taken as the systematic uncertainty.

The uncertainty due to FSR effect is checked with radiative Bhabha events. The data-MC difference of the efficiencies for the requirement of including FSR photons, 0.5\%, is taken as the systematic uncertainty. The uncertainty from the quoted BF for $\tau^+ \to e^+ \nu_e \bar{\nu}_{\tau}$ is 0.2\%~\cite{pdg}. The quadratic sum of all tag-mode independent systematic uncertainties is 2.2\%.

%------------------------
The last column of Table~\ref{dtyields_efficiency} lists the obtained $\mathcal{B}_{D_s^+\to\tau^+ \nu_{\tau}}$ for various tag modes. 
Weighting them by the inverse squares of the combined statistical and tag-mode dependent systematic uncertainties, we obtain $\mathcal{B}_{D_s^+ \to \tau^+ \nu_{\tau}}= (5.27\pm0.10\pm0.12)\times10^{-2}$. Here, the first error is statistical, and the second is the quadrature sum of the tag-mode dependent ($0.040\times10^{-2}$) and independent ($0.117\times10^{-2}$) systematic uncertainties. 
Combining Eq.~(\ref{decayratedstolnu_draft}) with $\mathcal{B}_{D_s^+ \to \tau^+ \nu_\tau} = \tau_{D_s^+} \Gamma_{D_s^+ \to \tau^+ \nu_\tau}$, where $\tau_{D_s^+}$ is the $D_s^+$ lifetime~\cite{pdg}, we find $f_{D_s^+} |V_{cs}| = (244.4\pm2.3\pm2.9)$ MeV. 
With $|V_{cs}|=0.97320\pm0.00011$ from the CKMfitter group~\cite{pdg}, we obtain $f_{D_s^+}=(251.1\pm2.4\pm3.0)$ MeV. 
Alternatively, taking $f_{D_s^+}=(249.9\pm0.5)$ MeV averaged from LQCD calculations~\cite{fds_flag}, we determine $|V_{cs}|=0.978\pm0.009\pm0.012$.

Based on our result of $f_{D_s^+}|V_{cs}|$ and the measured $f_{D^+}|V_{cd}|$ in Ref.~\cite{paper_fdvcd}, along with $|V_{cd}/V_{cs}| = 0.23259\pm0.00049$ from the SM global fit~\cite{pdg}, it yields $f_{D_s^+}/f_{D^+} = 1.244\pm0.017\pm0.021$, which is consistent with the LQCD calculation~\cite{fds_flag} within 2.4$\sigma$. Alternatively, taking $f_{D_s^+}/f_{D^+} = 1.1783\pm0.0016$ calculated by LQCD~\cite{fds_flag} as input, we extract $|V_{cd}/V_{cs}|^2 = 0.049\pm0.001\pm0.002$, which agrees with the value obtained from the CKM fitter within 2.3$\sigma$. 

Combining our measured BF, the world average value~\cite{pdg} of $\mathcal{B}_{D_s^+ \to \tau^+ \nu_\tau}$ can be improved to be $(5.34\pm0.13)\times 10^{-2}$. Using the world average of $\mathcal{B}_{D_s^+ \to \mu^+ \nu_\mu}$~\cite{pdg}, the ratio of the BFs is determined to be $\mathcal{B}_{D_s^+ \to \tau^+ \nu_{\tau}}/\mathcal{B}_{D_s^+ \to \mu^+ \nu_{\mu}}=9.72\pm0.37$, which is consistent with the SM prediction $9.75\pm 0.01$.

In summary, with data samples corresponding to an integrated luminosity of 6.32 fb$^{-1}$ collected at center-of-mass energies between 4178 and 4226 MeV, we present a precise measurement of the absolute BF for $D_s^+\to\tau^+ \nu_{\tau}$. The precision is improved by a factor of two compared to the previous best measurement~\cite{dstotaunu_hajime}. Taking inputs from the SM global CKM fit and LQCD separately, the decay constant $f_{D_s^+}$ and the magnitude of the CKM matrix element $|V_{cs}|$ are also extracted individually; all are the most precise results to date. Combining our result with the world average $\mathcal{B}_{D_s^+\to\tau^+ \nu_{\tau}}$ and $\mathcal{B}_{D_s^+\to\mu^+ \nu_{\mu}}$, we test the LFU in $\tau$-$\mu$ flavors, and no LFU violation is found with the current precision.

\section*{\boldmath ACKNOWLEDGMENTS}

The BESIII collaboration thanks the staff of BEPCII and the IHEP computing center for their strong support. This work is supported in part by National Key R\&D Program of China under Contracts Nos. 2020YFA0406400, 2020YFA0406300; National Natural Science Foundation of China (NSFC) under Contracts No. 12105077, No. 11805037, No. 11875054, No. 11625523, No. 11635010, No. 11735014, No. 11822506, No. 11835012, No. 11875122, No. 11935015, No. 11935016, No. 11935018, No. 11961141012, No. 12022510, No. 12025502, No. 12035009, No. 12035013, No. 12061131003; the Chinese Academy of Sciences (CAS) Large-Scale Scientific Facility Program; Joint Large-Scale Scientific Facility Funds of the NSFC and CAS under Contracts No. U1832121, No. U1732263, No. U1832207; CAS Key Research Program of Frontier Sciences under Contract No. QYZDJ-SSW-SLH040; 100 Talents Program of CAS; INPAC and Shanghai Key Laboratory for Particle Physics and Cosmology; Excellent Youth Foundation of Henan Province No. 212300410010; The youth talent support program of Henan Province No. ZYQR201912178; The Program for Innovative Research Team in University of Henan Province No. 19IRTSTHN018; ERC under Contract No. 758462; European Union Horizon 2020 research and innovation programme under Contract No. Marie Sklodowska-Curie grant agreement No 894790; German Research Foundation DFG under Contracts No. 443159800, Collaborative Research Center CRC 1044, FOR 2359, FOR 2359, GRK 214; Istituto Nazionale di Fisica Nucleare, Italy; Ministry of Development of Turkey under Contract No. DPT2006K-120470; National Science and Technology fund; Olle Engkvist Foundation under Contract No. 200-0605; STFC (United Kingdom); The Knut and Alice Wallenberg Foundation (Sweden) under Contract No. 2016.0157; The Royal Society, UK under Contracts No. DH140054 and No. DH160214; The Swedish Research Council; U. S. Department of Energy under Contracts No. DE-FG02-05ER41374, No. DE-SC-0012069.

\end{document}

%% file: author.tex
%\author{Author list}
\begin{small}
\begin{center}
M.~Ablikim$^{1}$, M.~N.~Achasov$^{10,c}$, P.~Adlarson$^{67}$, S. ~Ahmed$^{15}$, M.~Albrecht$^{4}$, R.~Aliberti$^{28}$, A.~Amoroso$^{66A,66C}$, M.~R.~An$^{32}$, Q.~An$^{63,49}$, X.~H.~Bai$^{57}$
% ...
\\
\vspace{0.2cm}
(BESIII Collaboration)\\
\vspace{0.2cm} {\it
$^{1}$ Institute of High Energy Physics, Beijing 100049, People's Republic of China\\
% ...
$^{4}$ Bochum Ruhr-University, D-44780 Bochum, Germany\\
% ...
$^{10}$ G.I. Budker Institute of Nuclear Physics SB RAS (BINP), Novosibirsk 630090, Russia\\
% ...
$^{15}$ Helmholtz Institute Mainz, Staudinger Weg 18, D-55099 Mainz, Germany\\
% ...
$^{28}$ Johannes Gutenberg University of Mainz, Johann-Joachim-Becher-Weg 45, D-55099 Mainz, Germany\\
% ...
$^{32}$ Liaoning Normal University, Dalian 116029, People's Republic of China\\
% ...
$^{49}$ State Key Laboratory of Particle Detection and Electronics, Beijing 100049, Hefei 230026, People's Republic of China\\
% ...
$^{57}$ University of Jinan, Jinan 250022, People's Republic of China\\
% ...
$^{63}$ University of Science and Technology of China, Hefei 230026, People's Republic of China\\
% ...
$^{66}$ University of Turin and INFN, (A)University of Turin, I-10125, Turin, Italy; (B)University of Eastern Piedmont, I-15121, Alessandria, Italy; (C)INFN, I-10125, Turin, Italy\\
$^{67}$ Uppsala University, Box 516, SE-75120 Uppsala, Sweden\\
% ...
$^{c}$ Also at the Novosibirsk State University, Novosibirsk, 630090, Russia\\
% ...
}\end{center}

\vspace{0.4cm}
\end{small}

%% file: dstotaunu.bbl
\begin{thebibliography}{**}
\bibitem{charge_conjugate} Throughout this Letter, charge conjugate channels are implied.
\bibitem{mass_window} The reason why we select different $M_{\rm rec}$ mass windows is to maintain roughly constant ST efficiencies for different values of $E_{\rm cm}$, while it is to suppress $D_s$ from $D_s^{*} D_s^{*}$ events for the data sample at $E_{\rm cm}=$ round 4226 MeV.
\bibitem{goodshower} Each good EMC shower has an energy of at least 25 MeV in the barrel region($|\cos \theta|<0.8$) or 50 MeV in the end-cap region ($0.86<|\cos \theta|<0.92$) of the EMC, where $\theta$ is the polar angle relative to the positron beam direction.
\end{thebibliography}
